\documentclass[aip,graphicx, reprint]{revtex4-2}

\draft % marks overfull lines with a black rule on the right

\usepackage{graphicx}% Include figure files
\usepackage{dcolumn}% Align table columns on decimal point
\usepackage{bm}% bold math
\usepackage[utf8]{inputenc}
\usepackage[T1]{fontenc}
\usepackage{mathptmx}
\usepackage{etoolbox}
\usepackage{color}

\usepackage{amsmath}
\makeatletter
\def\@email#1#2{%
 \endgroup
 \patchcmd{\titleblock@produce}
  {\frontmatter@RRAPformat}
  {\frontmatter@RRAPformat{\produce@RRAP{*#1\href{mailto:#2}{#2}}}\frontmatter@RRAPformat}
  {}{}
}%
\makeatother
\begin{document}

% Use the \preprint command to place your local institutional report number 
% on the title page in preprint mode.
% Multiple \preprint commands are allowed.
%\preprint{}

\title{$N$-qubit universal quantum logic with a photonic qudit and $O(N)$ linear optics elements} %Title of paper

% repeat the \author .. \affiliation  etc. as needed
% \email, \thanks, \homepage, \altaffiliation all apply to the current author.
% Explanatory text should go in the []'s, 
% actual e-mail address or url should go in the {}'s for \email and \homepage.
% Please use the appropriate macro for the type of information

% \affiliation command applies to all authors since the last \affiliation command. 
% The \affiliation command should follow the other information.

\author{Aymeric Delteil}
%\email[]{Your e-mail address}
%\homepage[]{Your web page}
%\thanks{}
%\altaffiliation{}
\affiliation{Universit\'e Paris-Saclay, UVSQ, CNRS,  GEMaC, 78000, Versailles, France. \\
aymeric.delteil@uvsq.fr}

% Collaboration name, if desired (requires use of superscriptaddress option in \documentclass). 
% \noaffiliation is required (may also be used with the \author command).
%\collaboration{}
%\noaffiliation

\date{\today}

\begin{abstract}
High-dimensional quantum units of information, or qudits, can carry more than one quantum bit of information in a single degree of freedom, and can therefore be used to boost the performance of quantum communication and quantum computation protocols. A photon in a superposition of $2^N$ time bins -- a time-bin qudit -- contains as much information as $N$ qubits. Here, we show that $N$-qubit states encoded in a single time-bin qudit can be arbitrarily and deterministically generated, manipulated and measured using a number of linear optics elements that scales linearly with $N$, as opposed with prior proposals of single-qudit implementation of $N$-qubit logic, which typically requires $O(2^N)$ elements. The simple and cost-effective implementation we propose can be used as a small-scale quantum processor. We then demonstrate a path towards scalability by interfacing distinct qudit processors to a matter qubit (atom or quantum dot spin) in an optical resonator. Such a cavity QED system allows for more advanced functionalities, such as single-qubit nondemolition measurement and two-qubit gates between distinct qudits. It could also enable quantum interfaces with other matter quantum nodes in the context of quantum networks and distributed quantum computing.
\end{abstract}

% insert suggested PACS numbers in braces on next line

\maketitle %\maketitle must follow title, authors, abstract and \pacs

\section{Introduction}

Quantum degrees of freedom with more than two orthogonal states, or qudits, have been proposed and used to improve the performance of  protocols in quantum computing~\cite{Wang20, Chi22, Kiktenko23}, quantum simulation~\cite{Neeley09} and quantum communication~\cite{Lee19, Tsuchimoto17}. Some photonic degrees of freedom, such as spatial mode (or path), orbital angular momentum, frequency and time bin, have a virtually unlimited number of states. It has therefore been proposed to encode $N$ logical qubits (described by a basis of $d = 2^N$ states of the quantum register) in $2^N$ states of the qudit by mapping the qubit states $|c_0, c_1\ldots c_N\rangle$ on the qudit states $|j\rangle, j = 0\ldots 2^{N-1}$ realized by spatial modes\cite{Cerf98} and orbital degrees of freedom~\cite{Garcia11}. In such proposals, however, at least some tasks require a number of optical elements of order $2^N$, making their implementation impractical for more than a few logical qubits -- in practice, experimental demonstrations have not exceeded three qubits~\cite{Fiorentino04, Kim07, Kagalwala17}. In the same spirit, other implementations of multiqubit systems in a single degree of freedom of dimension $d > 2$ (although often $d < 10$, \textit{i.e.} $N \leq 3$) have been also proposed using multilevel observables such as nuclear spins~\cite{Kessel99, Kessel02} or superconducting circuits~\cite{Kiktenko15a}, opening the way for a possible implementation of the Deutsch algorithm with a superconducting ququint ($d = 5$)~\cite{Kiktenko15}, and universal 3-qubit quantum logic using a trapped ion quoct ($d = 8$)~\cite{Nikolaeva24}. Remarkably, even protocols based on a single qudit of small dimension have the potential to provide quantum speedup~\cite{Gedik15,Zhan15}. However, in those cases, scaling up is precluded by the finite dimension of the qudit Hilbert space. On the other hand, despite the virtually unlimited dimension of the photon emission time (or time bin) Hilbert space, to the best of our knowledge, no such proposal has been based on photonic time-bin qudits.

Here, we extend previous proposals to time-bin qudits and we show that universal quantum logic can be implemented using $O(N)$ linear optics elements~\cite{Knill01, Kok07}, such as beamsplitters, phase shifters and electro-optic modulators, taking full benefit from the sequential nature of time bins. The approach for single- and two-qubit gates is fully deterministic and the gate fidelity is given by the fidelity of classical optics elements. It is moreover fully programmable and compatible with feed forward. While the implementation of $N$ qubits by a single qudit is not scalable \textit{per se}, we show that our approach can increase the number of logical qubits encoded in a single photon by at least one order of magnitude with current technology.

We then propose a path towards scalability based on the controlled interaction of the qudit with a matter qubit -- an atom or quantum dot in a cavity --, allowing for two-qubit quantum gates between distinct qudits~\cite{Hacker16}, as well as a number of potential other extensions such as single-qubit quantum nondemolition measurements. Other natural extensions could include a generalization of the architecture to multiple photonic qudits in the same circuit~\cite{Imany19}, thereby greatly extending the Hilbert space dimension implemented in photonic processors with respect to both their multiple-qubit and their single-qudit counterparts.

\section{A universal time-bin qudit quantum processor}

In this section, we describe the architecture allowing universal quantum computing with $N$ logical qubits encoded in a single time-bin photonic qudit. We consider the system operated in isolation; possible extensions and interconnections with external physical systems will be discussed in section~III.

\subsection{Representing $N$ logical qubits with a time-bin qudit}

Let us consider a $N$-qubit quantum register, with the index $i = 0\ldots N-1$ labelling individual qubits, and $\left\lbrace|0\rangle_i, |1\rangle_i\right\rbrace$ the computational basis of the $i$th qubit. The $N$-qubit register basis states are the $2^N$ states of the form $|c_0\rangle_0 \otimes \cdots \otimes |c_{N-1}\rangle_{N-1}$ (with $c_i = 0,1$). For $j = 0 \ldots 2^N - 1$ we define $C_i(j)$ such that $C_i(j) = 0$ if the $i$th binary digit of $j$ is $0$ and $C_i(j) =1$ if the $i$th binary digit is $1$. With this notation, the general state of the $N$-qubit register writes:

\begin{equation}
|\Psi\rangle = \sum_{j = 0}^{2^N-1} \alpha_j \bigotimes_{i=0}^{N-1}|C_i(j)\rangle_i 
\label{Psi}
\end{equation}

This notation allows an easy mapping of the $2^N$ states of the qubit register to the $2^N$ states $|\psi_j\rangle$ of a qudit through the identification $|\psi_j\rangle \equiv \bigotimes_{i=0}^{N-1}|C_i(j)\rangle_i $ such that $|\Psi\rangle = \sum_{j = 0}^{2^N-1}\alpha_j|\psi_j\rangle$.

\begin{figure}
 \includegraphics[width=0.98\linewidth]{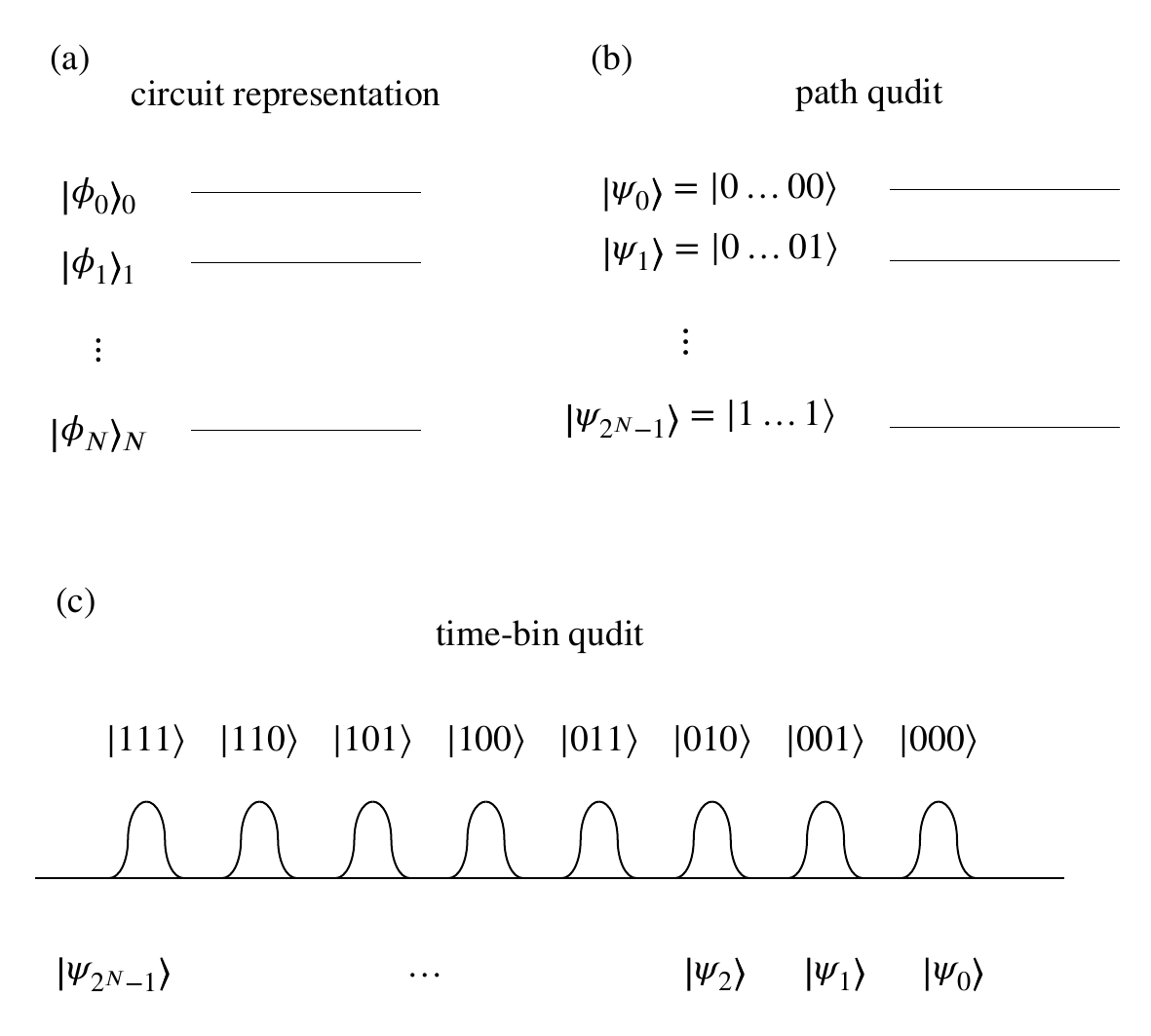}%
 \caption{\label{qubits} (a) $N$-qubit register in the circuit representation. (b) $2^N$ spatial modes can represent $N$ qubits. (c) time-bin representation of the $N$-qubit register.}%
 \end{figure}

The qudit states $|\psi_j\rangle$ can be implemented by high-dimensional degrees of freedom of photons, such as spatial mode (or path), frequency and time bin. Figure~\ref{qubits} shows the mapping of $N$ qubits to $2^N$ spatial modes (\textit{i.e.} a path qudit) as well as $2^N$ time bins. The general state $|\Psi\rangle$ is a linear superposition of these degrees of freedom.

\subsection{Processor architecture}

In this section, we describe the general architecture of an optical setup emulating a $N$-qubit quantum processor based on a photonic qudit. The proposed implementation of single- and two-qubit gates will be detailed in the following sections.

\begin{figure}
 \includegraphics[width=0.98\linewidth]{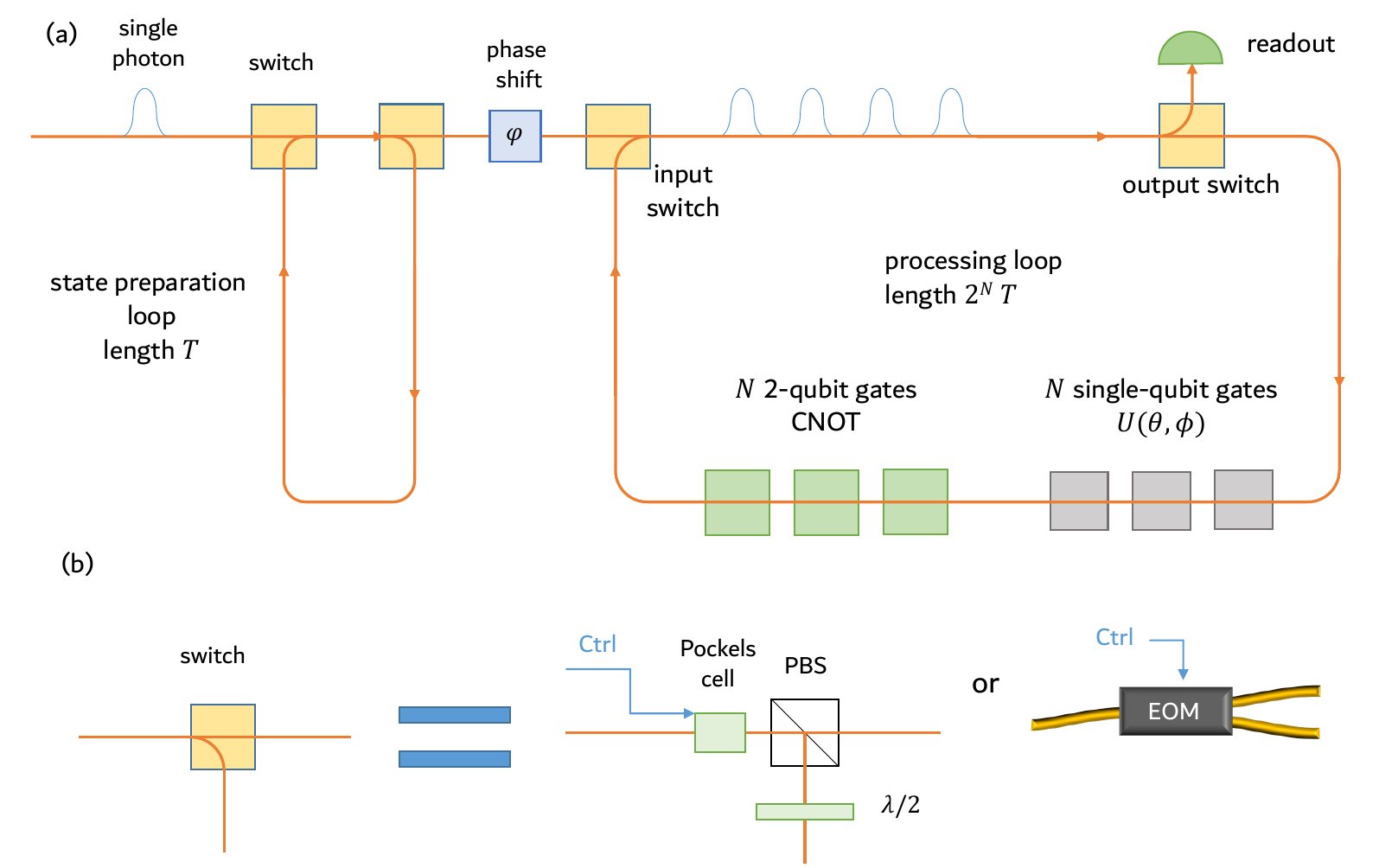}%
 \caption{\label{architecture} (a) Scheme of the qudit processor architecture. (b) Possible implementation of the photonic switches. The proposed implementations of single- and two-qubit gates (resp. gray and green squares) are detailed in figures~\ref{1qubit_implementation} and~\ref{2qubit_implementation}.}%
 \end{figure}

Figure~\ref{architecture}a depicts the processor. It is based on a loop architecture, in a similar way to prior work on boson sampling~\cite{He17}. The main loop is referred to as the processing loop. Its optical length, together with the time $T$ between two time bins, determines the number of bins -- and therefore the number $N$ of logical qubits implemented by the system. In section~IV, we discuss realistic numbers for $T$ and $N$. Fast switches are used to inject the initial state in the processing loop, as well as to extract the final state for readout. Two examples of implementation of such fast switches are shown Figure~\ref{architecture}b, based on either a Pockels cell and a polarizing beam splitter, which is common for deterministic time-to-path demultiplexing in photonic quantum computing~\cite{Wang19, Maring24}, or a dual-output fiber electro-optic modulator, which can be used in a fully fiber-based approach. Universal computing is performed by $N$ single-qubit gates and $N$ two-qubit gates, which we justify in the following sections.

For the sake of generality, we propose a way to prepare any arbitrary initial states of the qudit (and therefore of the $N$ logical qubits), although in principle the preparation of only one given computational state is sufficient. The general state preparation can be performed based on an upstream loop whose length defines the repetition time $T$. A single-photon pulse is injected in the loop using an input switch. A second switch defines at each pass which amplitude is transmitted (and therefore extracted from the preparation loop). An additional phase shifter can be set to control the phase of the time bin. The transmitted component after a number of loops $j$ defines the complex amplitude $\alpha_j$. In this way, any state $|\Psi\rangle$ can be generated. Alternative ways to prepare arbitrary time-bin qudit states can be found in the literature~\cite{Zheng22}. We note that preparation of the computational basis state $|\Psi\rangle = |\psi_0\rangle$ is trivial and does not require the preparation loop -- but only that a single-photon pulse is injected in the processing loop. For more complex implementations, qudits entangled with external degrees of freedom can also be injected in the processing loop.

Readout is trivial in the computational basis. The output switch is set to the readout port and a time-correlated single-photon counter (TCSPC) projects the qudit in one of the $2^N$ computational basis states $|\psi_j\rangle$, which is equivalent to a full readout of the $N$-qubit register. Together with the universal set of quantum gates we propose in this work, this allows to measure the register in any basis. Alternatively, other qudit measurement techniques can be used, allowing full quantum state tomography~\cite{Ikuta17}. Moreover, in section~III we propose a solution for partial measurement of subsets of logical qubits.

\subsection{Single-qubit gates}

In order to build single-qubit gates in the qudit representation, it is useful to define additional notations. We note $Z_i$ the set of integers $j$ verifying $j \in \{0 \ldots 2^N-1\}$ and such that the $i$th binary digit of $j$ (in base~2) is zero, \textit{i.e.} $C_i(j) = 0$. For instance, $Z_0 = \{0, 2, 4 \ldots \}$, $Z_1 = \{0, 1, 4, 5, 8, 9 \ldots \}$. We note $\overline{Z_i}$ the complementary set (which has the same cardinality). Some noteworthy properties follow. First, for a given $i$, if $j \in Z_i$ then $j + 2^i \in \overline{Z_i}$. In their binary representation, $j$ and $j + 2^i$ only differ by the value of the $i$th bit. Therefore, if $j \in Z_i$, $|\psi_j\rangle$ and $|\psi_{j + 2^i}\rangle$ represent two states differing only by the state of the $i$th qubit, with $|C_i(j)\rangle_i = |0\rangle_i$ and $|C_i(j + 2^i)\rangle_i = |1\rangle_i$.

Consequently, the general state $|\Psi\rangle$ described by equation~\ref{Psi} can be rewritten in the following way, where the qubit $i$ is singled out:

\begin{equation}
|\Psi\rangle = \sum_{j \in Z_i} \bigotimes_{k\neq i}|C_k(j)\rangle_k \otimes \left(
\alpha_j |0\rangle_i + \alpha_{j+2^i}|1\rangle_i
\right) 
\label{Psi_i}
\end{equation}

A single-qubit gate $U$ acting on the $i$th qubit can be written as $U_i = \bigotimes_{k\neq i}{I_k}\otimes U$. Applying this operator on $|\Psi\rangle$ therefore yields

\begin{align}
\nonumber U_i |\Psi\rangle &= \sum_{j = 0}^{2^N-1} \alpha_j \bigotimes_{k \neq i}|C_k(j)\rangle_k  \otimes U |C_i(j)\rangle_i \\ 
~&= \sum_{j \in Z_i} \bigotimes_{k\neq i}|C_k(j)\rangle_k \otimes U \left(
\alpha_j |0\rangle_i + \alpha_{j+2^i}|1\rangle_i
\right) 
\label{Upsi1}
\end{align}

where we used the fact that $|C_k(j)\rangle = |C_k(j + 2^i)\rangle$ for $j \in Z_i$ and $k\neq i$. Finally, equation~\ref{Upsi1} can be rewritten

\begin{equation}
U_i |\Psi\rangle = \sum_{j \in Z_i} U(\alpha_j |\psi_j\rangle +  \alpha_{j+2^i}|\psi_{j + 2^i}\rangle)
\label{Upsi2}
\end{equation}

\begin{figure}
 \includegraphics[width=0.98\linewidth]{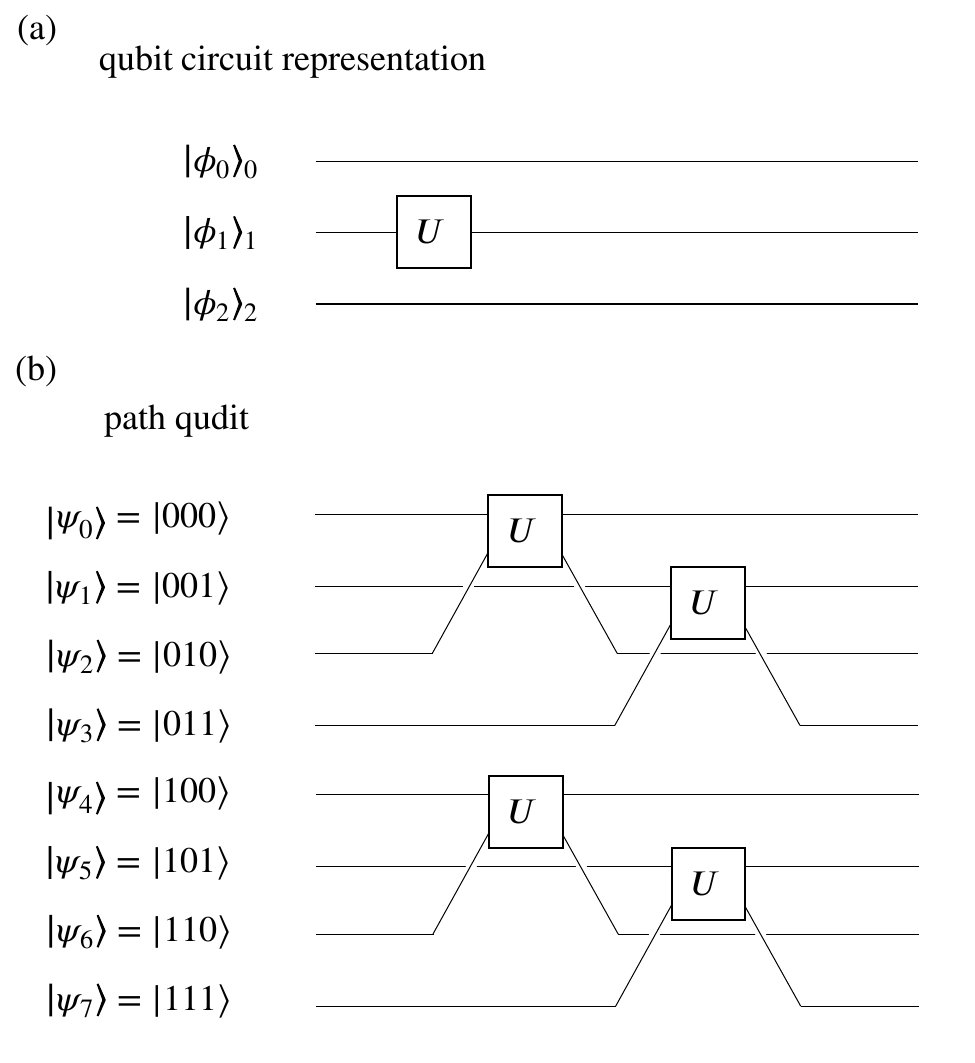}%
 \caption{\label{1qubit_gate} (a) Circuit representation of a single-qubit gate applied to the qubit $i = 1$ of a 3-qubit register. (b) Equivalent operation in a path qudit: the gate is applied to all pairs of modes whose label $j$ differ by the value of the $i$th digit.}%
 \end{figure}

Therefore, a single-qubit gate $U$ on the logical qubit $i$ is equivalent to applying the same gate $U$ to all pairs of states $j$ and $j+2^i$. This requires $2^{N-1}$ operations. Figure~\ref{1qubit_gate} illustrates how this would be implemented to a spatial mode qudit -- in which case such operation would be rapidly intractable as $N$ grows.

\begin{figure}
  \includegraphics[width=\linewidth]{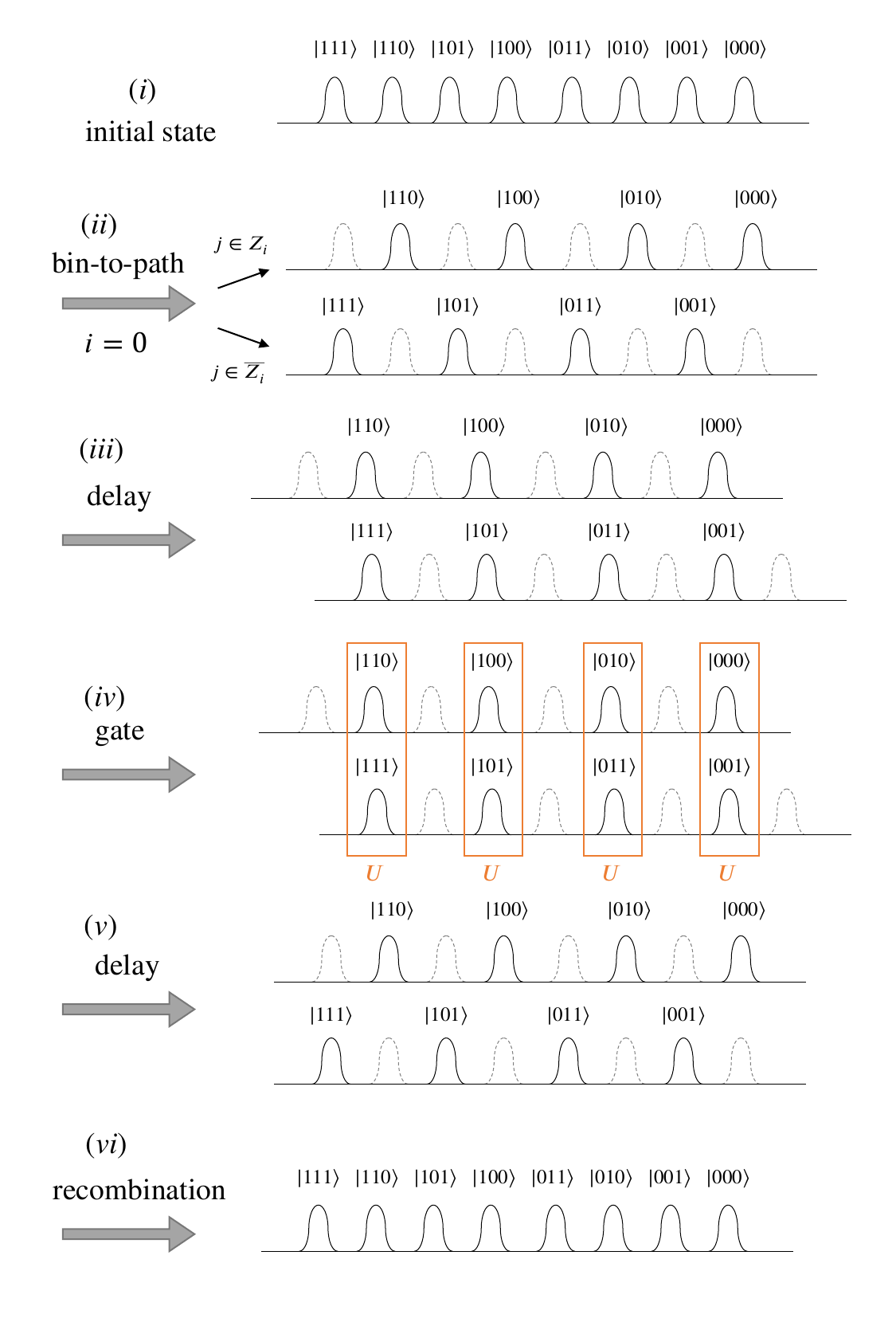}%
 \caption{\label{1qubit_gate_timebin} Principle of a single-qubit gate applied to qubit~0. (\textit{i}) Initial state $|\Psi\rangle$, in a superposition of $d$ time bins (the coefficients $\alpha_j$ are not depicted). (\textit{ii}) The bins are separated according to $C_0(j)$, yielding a parity-wise separation. (\textit{iii}) The odd bins are delayed by $T$ such that all pairs of bins differing only by the value of $C_0(j)$ are simultaneous. (\textit{iv}) The gate $U$ is then applied to all pairs of simultaneous bins. (\textit{v}) One of the output modes is then delayed by $T$. (\textit{vi}) Finally, the bins are interleaved back.}%
 \end{figure}

In the case of time-bin qudits however, this set of operations can be done sequentially, based on the observation that all pairs of states (or time bins) to which $U$ is applied are separated by the same time delay $2^iT$. The principle, described in figure~\ref{1qubit_gate_timebin}, is then to separate the bins $\{|\psi_j\rangle\}$ into two spatial modes conditioned on $C_i(j)$ (i.e. the binary value of the $i$th binary digit of $j$), to then delay one of the two modes by $2^i T$, to sequentially apply the same single-qubit gate to all pairs of bins, to delay again one of the arms by $2^i T$ and to finally recombine the two modes.

\begin{figure}
  \includegraphics[width=\linewidth]{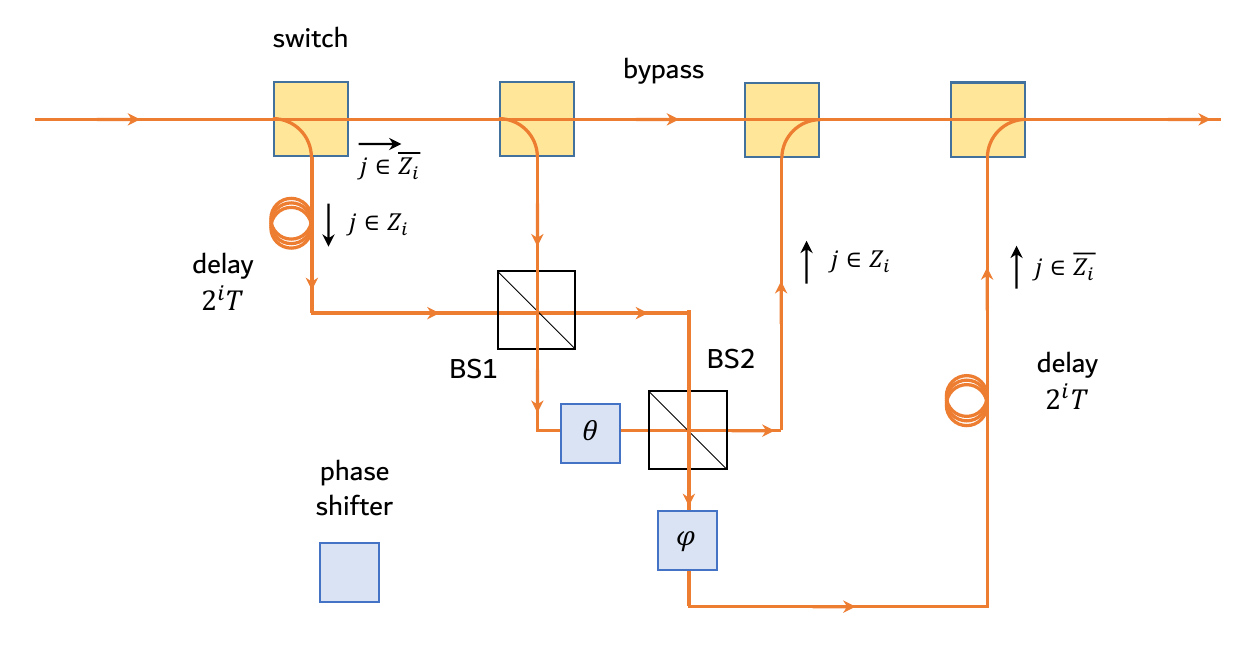}%
 \caption{\label{1qubit_implementation} Implementation of a single-qubit gate on qubit $i$ switches, fiber delays and phase shifters. The labels H and V denote the polarization state. The topmost horizontal mode is the processing loop in which the single-qubit gates are inserted, as depicted by the gray boxes on figure~\ref{architecture}a.}%
 \end{figure}

An example of implementation is given figure~\ref{1qubit_implementation}. It is based on a universal optical quantum gate~\cite{Reck94} using a Mach-Zehnder interferometer (with beamsplitters labeled BS1 and BS2) and two phase shifters ($\theta$ and $\varphi$). Two switches separate the bins $j$ according to the value of $C_i(j)$ and a delay line allows the bins $j$ and $j+2^i$ to arrive simultaneously at the interferometer input ports. One of the output ports is also delayed by the same amount before the bins from the two output ports are interleaved. The gate can also be bypassed in case it is not needed. The single-qubit gate acting on qubit $i$ differs from the other single-qubit gates only by the length of the two delay lines. Thus, $N$ different single-qubit gates can be inserted in the processing loop (as shown by the gray boxes on figure~\ref{architecture}a) for complete control of all single logical qubits.

\subsection{Two-qubit gates}

Universal quantum logic also requires a two-qubit gate. Here, we show how a CNOT gate can be applied to any pair of qubits $m$ and $n$, where $m$ ($n$) labels the control (target) qubit. In the same spirit as for the single-qubit gates, let us first rewrite the general state $ |\Psi\rangle$ in a way that singles out the qubits $m$ and~$n$:

\begin{align}
\nonumber |\Psi\rangle = \sum_{j \in Z_m \cap Z_n} \bigotimes_{k\neq m,n}
&|C_k(j)\rangle_k \otimes (
\alpha_j |0_m, 0_n\rangle + \alpha_{j+2^n}|0_m, 1_n\rangle + \\ ~&
 \alpha_{j+2^m}|1_m, 0_n\rangle +  \alpha_{j+2^m + 2^n}|1_m, 1_n\rangle
) 
\label{Psi_mn}
\end{align}

such that the effect of the CNOT gate on qubits $m$ and $n$ can be written

\begin{align}
\nonumber \mathrm{CNOT}_{m,n}|\Psi\rangle &= \sum_{j \in Z_m \cap Z_n} \bigotimes_{k\neq m,n}
|C_k(j)\rangle_k \otimes (
\alpha_j |0_m, 0_n\rangle +  \\ \nonumber ~&
\alpha_{j+2^n}|0_m, 1_n\rangle + \alpha_{j+2^m}|1_m, 1_n\rangle +  \alpha_{j+2^m + 2^n}|1_m, 0_n\rangle ) \\
~& = \sum_{j \in Z_m } \alpha_j |\psi_j\rangle + \sum_{j \in \overline{Z_m} \cap Z_n } \left(\alpha_{j + 2^n} |\psi_j\rangle + \alpha_j |\psi_{j + 2^n}\rangle \right)
\label{CNOT_mn}
\end{align}

\begin{figure}
  \includegraphics[width=0.8\linewidth]{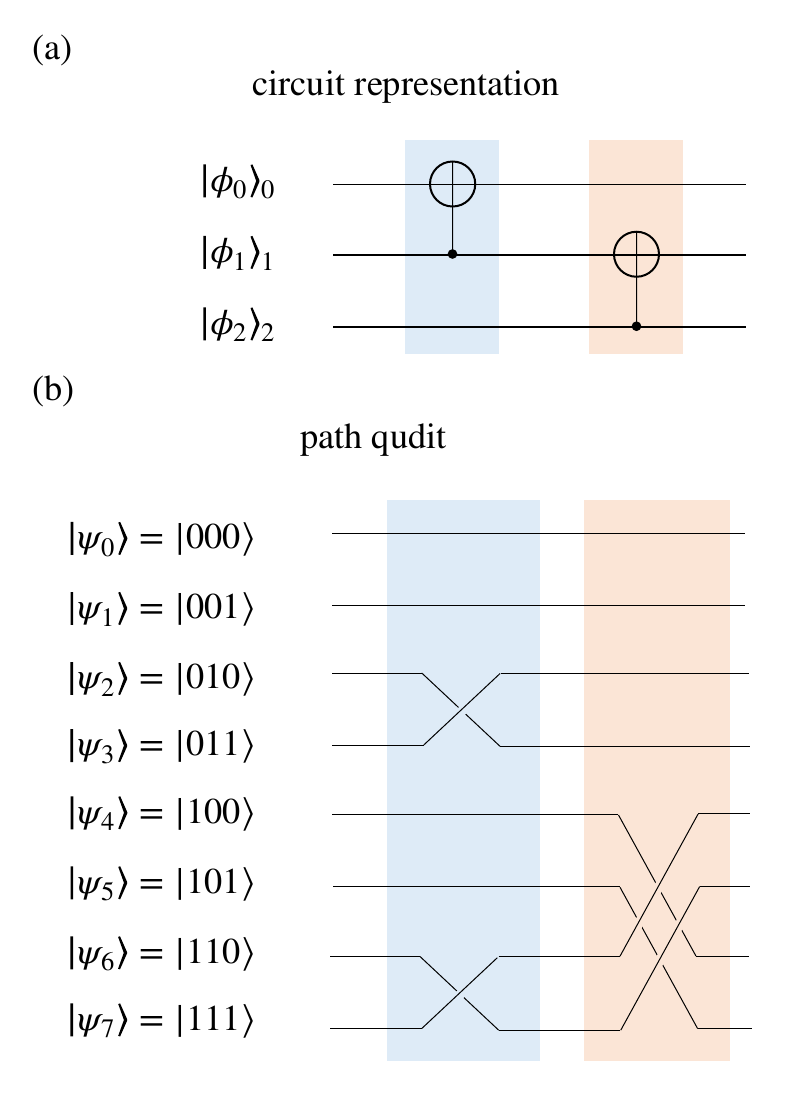}%
 \caption{\label{2qubit_gate} Two examples of CNOT gates in the circuit representation (a) and in a spatial mode qudit representation (b). In the first (second) example, qubit~1 (qubit~2) is the control qubit and qubit~0 (qubit~1) is the target qubit.}%
 \end{figure}

The effect of a CNOT gate on qubits $m$ and $n$ in the qudit basis is therefore to swap all components separated by $2^n$ if $C_m(j) = 1$ (\textit{i.e.} for $j \in \overline{Z_m}$). In other words, the bins differing only by the value of the target qubit $n$ are swapped if the value of the control qubit $m$ equals 1. This corresponds to swapping $2^{N-1}$ bins separated by the same delay $2^n$ given by the choice of the target qubit. This operation is depicted in the circuit representation and in the spatial mode qudit representation on figure~\ref{2qubit_gate}. 

\begin{figure}
  \includegraphics[width=\linewidth]{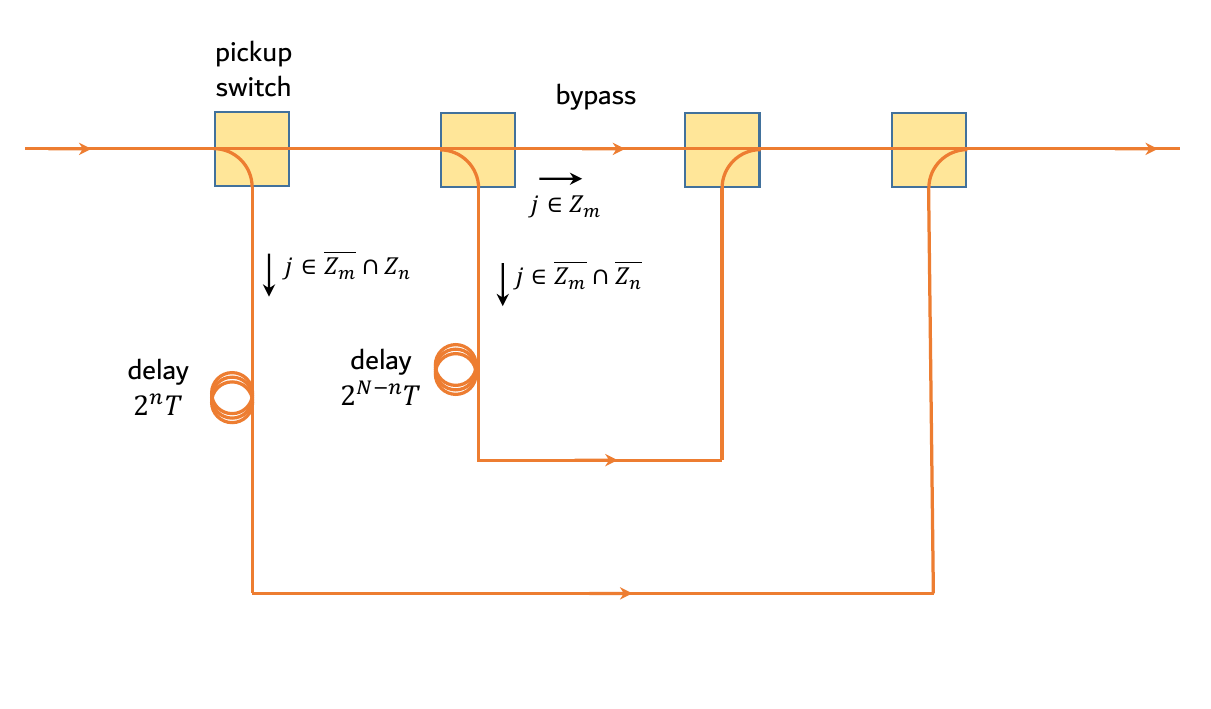}%
 \caption{\label{2qubit_implementation} A possible implementation of the CNOT gate. The delays set the choice of the target qubit~$n$. The choice of the control qubit $m$ is given by the electronics that control the switches. Therefore, this setup realizes all CNOT gates acting on qubit $n$ as the target qubit. The labels H and V denote the polarization state. The topmost horizontal mode is the processing loop in which the CNOT gates are inserted, as depicted by the green boxes on figure~\ref{architecture}a.}
 \end{figure}

To illustrate this, on figure~\ref{2qubit_implementation} we propose an implementation of the CNOT gate, based only on switches and fiber delays. The time bins $j$ such that $C_m(j)=1$ and $C_n(j)=0$ (resp. $C_m(j)=1$ and $C_n(j)=1$) are channelled to the delay line $2^n T$ (resp. $2^{N-n}T$), which swaps these two sets of bins after recombination. All other time bins are bypassed.

Remarkably, despite the fact that there are $N(N-1)$ different CNOT gates, there are only $N$ different optical delays to implement -- one for each target qubit. Indeed, given a target qubit (fixed by the delay lines on figure~\ref{2qubit_implementation}), any of the $N - 1$ possible control qubits can be selected electronically by the timing of the pulses sent to the switches. Therefore, the setup shown figure~\ref{2qubit_implementation} realizes all CNOT$_{m,n}$ for a fixed $n$, and there are only $N$ different CNOT blocks to implement, as depicted on figure~\ref{architecture}a (where each of the $N$ green boxes represent a CNOT setup as detailed on figure~\ref{2qubit_implementation}, with a fixed pair of delays). This is a specific advantage offered by the time bin encoding.

\section{Cavity QED for single-qubit QND measurements and two-qubit gates from distinct qudits}

Cavity quantum electrodynamics (QED) systems have been proposed to mediate photon-photon interaction in a deterministic way~\cite{Duan04, Hu08}. Such systems can be realized using a single atom in a cavity~\cite{Hacker16}, or a solid-state quantum emitter in the strong or weak coupling regime~\cite{Kim13,Mehdi24}. Similar results can be obtained using waveguide QED~\cite{Appel22, Chan23}. All these approaches are based on spin-dependent reflection coefficients.

\subsection{QND measurement of a single logical qubit}

Let us consider a cavity QED system with two spin ground states $|\uparrow\rangle$ and $|\downarrow\rangle$ -- say, a charged quantum dot (QD) -- in a cavity. The cavity and QD parameters can be set such that an incident photon will experience a different phase shift upon reflection on the cavity depending on the electron (or hole) spin state. Without loss of generality (owing to the possibility of performing local unitary operations on both the spin and the photon state depending on the particular implementation), we consider the reflection coefficient to be $1$ if the spin state is $|\uparrow\rangle$, and $-1$ if the spin state is $|\downarrow\rangle$~\cite{Duan04}. Therefore, by preparing the spin state in $|+\rangle = (|\uparrow\rangle + |\downarrow\rangle )/\sqrt{2} $, a photon number superposition $\alpha|0\rangle + \beta|1\rangle$ incident on the cavity leaves the spin-photon state in $\alpha|0\rangle|+\rangle + \beta|1\rangle|-\rangle$.

\begin{figure}
  \includegraphics[width=\linewidth]{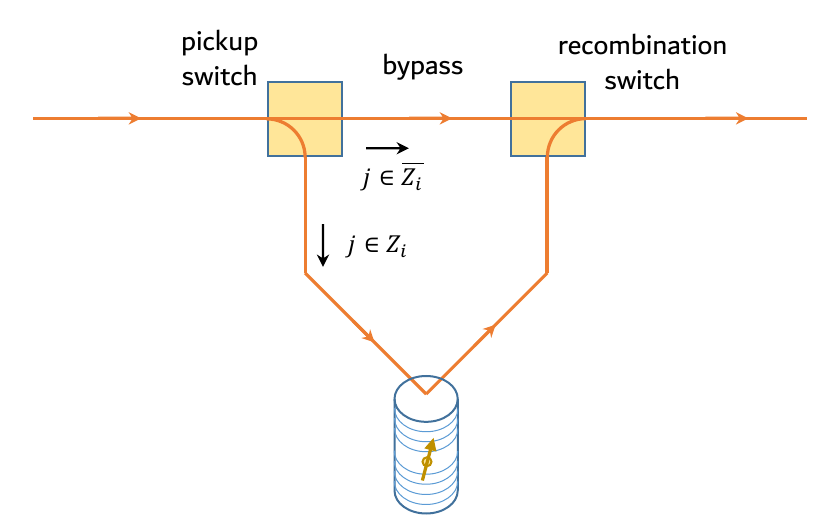}%
 \caption{\label{cQED_1qudit} Selective reflection of time bins on a cQED system: after spin preparation, the bins are channelled to the cavity depending on their label, so to entangle a single logical qubit to the cavity. Subsequent measurement of the spin state allows QND measurement of any given logical qubit in any basis.
}
 \end{figure}

We now consider the setup shown figure~\ref{cQED_1qudit}, which can send any set of bins of a qudit to make them reflect on the cavity. Based on eq.~\ref{Psi_i}, if we initialize the spin in $|+\rangle$, and then channel to the cavity the bins $j \in \overline{Z_i}$, we obtain the resulting spin-photon entangled state:

\begin{equation}
\sum_{j \in Z_i} \bigotimes_{k\neq i}|C_k(j)\rangle_k \otimes \left(
\alpha_j |0\rangle_i |+\rangle + \alpha_{j+2^i}|1\rangle_i |-\rangle
\right) 
\label{Psi_ispin}
\end{equation}

This expression shows that subsequent readout of the spin state in the $|+\rangle / |-\rangle$ rotated basis performs a quantum nondemolition measurement of the logical qubit $i$ in the computational basis. Alternatively, it can be measured in any other basis by choosing the appropriate measurement basis for the spin state. This ability to selectively measure any logical qubit in the register is useful for protocols such as quantum teleportation, algorithms based on quantum feedback such as quantum error correction codes, or one-way quantum computing based on cluster states. 

\subsection{Entangling gates between logical qubits in dinstinct qudits}

\begin{figure}
  \includegraphics[width=\linewidth]{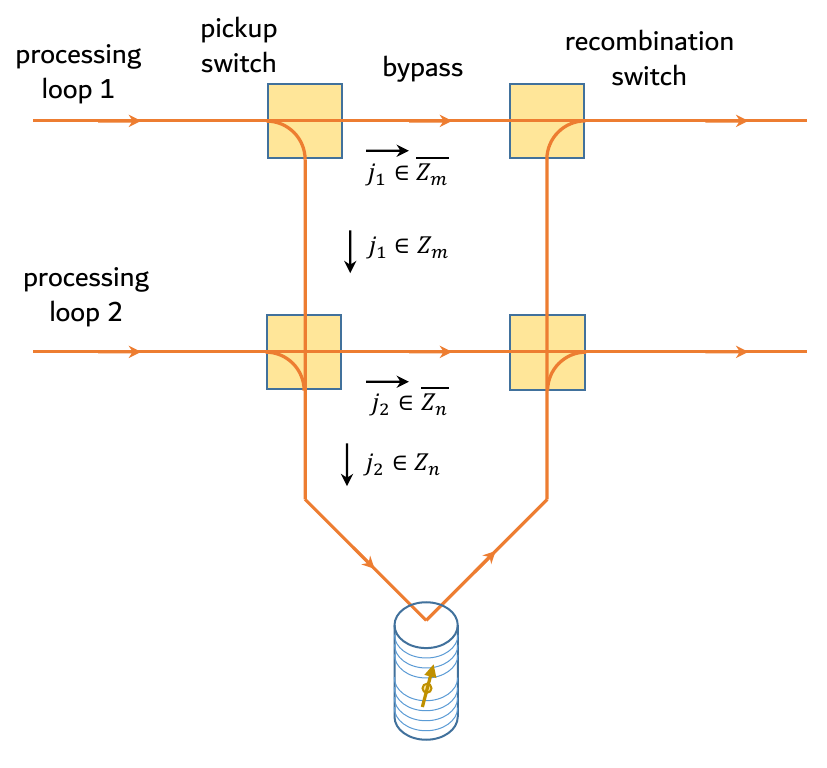}%
 \caption{\label{cQED_2qudit}  Selective reflection of time bins from different qudit circuits on a cQED system: after spin preparation, bins are channelled to the cavity conditioned on their qudit circuit and their label, so to entangle two single logical qubits with the cavity. Subsequent measurement of the spin state implements a two-qubit gate. }
 \end{figure}

The previous setup can be generalized to allow for the reflection of time bins from distinct qudits. This opens the way to two-qubit gates between logical qubits encoded in two different time-bin qudits. Figure~\ref{cQED_2qudit} provides a suggestion about how to implement such gates using a cQED system identical to the one considered in Sec.~III~A. 

We give the example of a CZ gate (or controlled phase flip gate), which is equivalent to a CNOT gate upon local unitary transformations. We consider a CZ gate between qubit $m$ of qudit~1 and qubit $n$ of qudit~2. We start from a spin prepared in the $|+\rangle$ state, and we apply the spin-photon mapping process described in Sec.~III~A. Upon reflection of bins $j_1 \in \overline{Z_m}$ of qudit~1, the state $|\Psi\rangle$ of the spin and the two qudits writes

\begin{align}
\nonumber
\sum_{j_1 \in Z_m} \bigotimes_{k_1\neq m}|C_{k_1}(j_1)\rangle_{k_1} \otimes \left(
\alpha_{j_1} |0\rangle_m |+\rangle + \alpha_{{j_1}+2^m}|1\rangle_m |-\rangle
\right) \\
\otimes \sum_{j_2 \in Z_n} \bigotimes_{{k_2}\neq n}|C_{k_2}(j_2)\rangle_{k_2} \otimes \left(
\alpha_{j_2} |0\rangle_n + \alpha_{j_2+2^n}|1\rangle_n 
\right)
\label{Psi_j1j2spin}
\end{align}

We then perform a $\pi/2$ rotation of the spin state to map $|+\rangle$ to $|\uparrow \rangle$ and $|-\rangle$ to $| \downarrow\rangle$. In the transformed state, the spin is $|\downarrow\rangle$ only when qubit $m$ is in the $|1\rangle$ state. Then, upon reflection of bins $j_2 \in \overline{Z_n}$ of the second qudit, a phase flip occurs if the spin is $\downarrow$ and the qubit $n$ of the second qudit is in the $|1\rangle$ state:

\begin{align}
\nonumber
 \sum_{j_1 \in Z_m} &\bigotimes_{k_1\neq m}|C_{k_1}(j_1)\rangle_{k_1} 
\otimes \sum_{j_2 \in Z_n} \bigotimes_{{k_2}\neq n}|C_{k_2}(j_2)\rangle_{k_2}
\\ \nonumber
\otimes & (
\alpha_{j_1} \alpha_{j_2} |0\rangle_m|0\rangle_n  |\uparrow \rangle
+ \alpha_{j_1} \alpha_{j_2 + 2^n} |0\rangle_m|1\rangle_n  |\uparrow \rangle
\\
&+ \alpha_{j_1 + 2^m} \alpha_{j_2} |1\rangle_m|0\rangle_n  |\downarrow \rangle
- \alpha_{j_1 + 2^m} \alpha_{j_2 + 2^n} |1\rangle_m|1\rangle_n  |\downarrow \rangle
) 
\label{Psi_j1j2spin2}
\end{align}

The next step is a second spin rotation by $-\pi/2$ followed by spin measurement in the computational basis. Detection of the spin in the $|\uparrow\rangle$ state disentangles it from the two-qudit state and projects it onto

\begin{align}
\nonumber
 \sum_{j_1 \in Z_m} &\bigotimes_{k_1\neq m}|C_{k_1}(j_1)\rangle_{k_1} 
\otimes \sum_{j_2 \in Z_n} \bigotimes_{{k_2}\neq n}|C_{k_2}(j_2)\rangle_{k_2}
\\ \nonumber
\otimes &(
\alpha_{j_1} \alpha_{j_2} |0\rangle_m|0\rangle_n  
+ \alpha_{j_1} \alpha_{j_2 + 2^n} |0\rangle_m|1\rangle_n 
\\
&+ \alpha_{j_1 + 2^m} \alpha_{j_2} |1\rangle_m|0\rangle_n 
- \alpha_{j_1 + 2^m} \alpha_{j_2 + 2^n} |1\rangle_m|1\rangle_n  
) 
\label{Psi_j1j2_cz}
\end{align}

which realizes the CZ gate. If the spin measurement outcome is $|\downarrow\rangle$, the resulting state differs from eq.~\ref{Psi_j1j2_cz} only by a local $Z$ gate on qubit~$m$.

This architecture can be generalized to implement the coupling of any pair of logical qubits encoded in $M$ qudits using the same cavity. Such generalized architecture implements a $M\times N$ qubits universal quantum computer with $O(M\times N)$ overhead and a single cQED system. Multiqubit entangled states such as cluster states or GHZ states can also directly be generated upon multi-qudit reflections separated by appropriate spin rotations. The spin-photon interface also allows to consider interconnections with quantum memories (such as nuclear spins) and interfaces with matter qubit processors for hybrid quantum systems.

\section{Discussion}

The set of single- and two-qubit gates exposed in section~II, although universal, is not exhaustive, and could be completed by other quantum gates built in the same spirit. For instance, a Toffoli gate with control qubits $m$ and $n$, and target qubit $p$ can be realized using the CNOT circuit by swapping bins $j$ and $j+2^p$ such that $j\in \overline{Z_m}\cap \overline{Z_n}\cap Z_p$. Generalized control gates up to $N$ qubits are similarly trivial to construct. Conversely, any gate demonstrated for time-bin qubits -- and for dual rail qubits -- could be adapted to the logical qubits represented by a time-bin qudit, such as controlled-phase gate~\cite{Lo18}. Since it is based on single-photon interference, experimental implementation will need phase stabilization, which requires overhead that also scales linearly with $N$, and can potentially be optimized to $O(\log(N))$ by distributing the reference phase.

For realistic implementations, current technology allows switching times, photon detection resolution, and single photon generation to be performed faster than 100~ps~\cite{Bull04}. With a conservative value for $T$ of 1~ns, and given that quantum protocols based on time-bin qubits have been performed with high fidelity using fiber lengths of more than 100~kilometers~\cite{Krutyanskiy24, Zhou24}, one can accomodate more than $10^6$ time bins, which implement 20~qubits. The scheme would be difficult to scale up further as such, due to the exponential cost of additional qubits on the fiber length and the associated exponential decrease in the clock rate. We expect that an implementation with 10-15~qubits would be relatively easy and low-cost to implement in a laboratory. This number of qubits is already comparable to the current state of the art of photonic quantum computers based on $N$ single-photon qubits. In the closed-loop version (\textit{i.e.} with a single qudit and no quantum interface, as in figure~\ref{architecture}), this architecture could serve as a few-qubit demonstrator for proofs of concepts, quantum simulation, fundamental tests of quantum mechanics~\cite{Lim14, Sadana22} or as a reference for quantum benchmarking of photonic quantum computers. 

Given that the operation only relies on single-photon interference and single-photon detection, and does not involve (effective) photon-photon interaction, there is no particular requirement in terms of purity and indistinguishability, as already pointed out in prior work on photonic qudits~\cite{Cerf98, Kwiat00}. This consideration allows to use multiphoton pulses as inputs, which can be seen as parallel attempts, with a single-photon detection heralding success~\cite{Kwiat00}. Therefore, the photon number in the input pulse can in principle be tuned in such a way to obtain an output photon detection probability close to unity, allowing to compensate for losses. This is a specific advantage of single qudits, allowed by the fact that entanglement is purely intra-particle and no coincidence is required, as opposed to more usual photonic quantum computing, where true single photons are needed and operations are performed through (effective) photon-photon interactions -- thus making purity and indistinguishability crucial. The success rate can even be pushed further provided that the detector is able to detect more than one photon. This can be done using a (generalized) Hanbury Brown and Twiss setup, or a photon-number resolving detector, and allows parallel runs emulating simultaneous operation of multiple identical processors.

The proposed extensions using cavity QED allow multiqubit operations on the logical qubits encoded in different qudit circuits, yielding potentially large scale computers. The interconnection of logical qubits from distinct qudits would require true single photons of high purity -- which could in principle be generated by the same cQED system used for two-qubit gates. On the other hand, indistinguishability is not strictly required~\cite{Hu08}, which could be an advantage compared to more traditional approaches for single-photon quantum computing. A $M \times N$ implementation would be sensitive to photon losses as $\eta^M$, where $\eta$ is the transmission of a single circuit. Success of the computation is heralded by detection of $M$ photons at the end of the computation. Further extensions based on interfaces to multiple matter qubits can be envisaged. Indeed, a single $2^N$-dimensional time-bin qudit can be interfaced with $N$-qubit quantum memories~\cite{Zheng22}, so to distribute entanglement between $N$ pairs of atoms, with an exponential reduction of the photon losses as compared with using $N$ photons as in more traditional cascaded quantum systems~\cite{Gardiner93, Carmichael93, Delteil17}. The possibility to operate quantum gates directly on the time-bin qudits could greatly expand the dimension of distant entangled states that can be generated. Future extensions can also include the integration of multiple qudits in the same loop circuit to expand the dimension of the Hilbert space using hyperentanglement~\cite{Deng17,Graffitti22}. This could be crucial to scale up the number of logical qubits in single-photon quantum computing, which is so far limited by the number of photons which can be demultiplexed from a single quantum emitter~\cite{Wang19,Maring24}.

\section{Conclusion}

 We have demonstrated that universal quantum logic, including state preparation, readout, single-qubit gates and CNOT gates between any pair of logical qubits, can be implemented for a time-bin qudit encoding $N$ qubits. This set of single- and two-qubit gates is fully programmable and only requires overhead of the order of $N$, which makes experimental implementation for a moderate number of qubits more realistic and, incidentally, should also yield a reduced error rate as compared with prior single-qudit architectures. We have then shown that an interface with matter qubits is possible using cavity QED systems, allowing further functionalities as well as interconnections between multiple qudits for further scalability, and opening the way to the integration into quantum networks. Finally, we have discussed the limits of technical implementations with current technologies, as well as other possible extensions of our architecture. Altogether, time-bin qudits are particularly well suited for encoding and manipulating an underlying collection of logical qubits, as well as for further integration into larger architectures.

\begin{acknowledgments}
We acknowledge fruitful discussions with J.-P. Hermier and S. Buil.
\end{acknowledgments}

% Create the reference section using BibTeX:
%\bibliography{your-bib-file}
~\\
%
%
%\section*{Data Availability Statement}
%The data that support the findings of this study are openly available in
%[repository name] at http://doi.org/[doi], reference number [reference number].

\end{document}